\documentclass[10pt]{article}

\usepackage{amsmath}
\usepackage{amssymb}

\usepackage{graphicx, subfigure}

\usepackage{cite}

\usepackage{color} 

\usepackage[labelfont=bf,labelsep=period,justification=raggedright]{caption}

\makeatletter
\renewcommand{\@biblabel}[1]{\quad#1.}
\makeatother

\date{}

\begin{document}

\begin{flushleft}
{\Large
\textbf{Improved haplotyping of rare variants using next-generation sequence data}
}
\\
Fouad Zakharia$^{1}$, 
Carlos D Bustamante$^{1, \ast}$, 
\\
\bf{1} Department of Genetics, Stanford University, Stanford, CA 94305, USA
\\
$\ast$ E-mail: zakharia@stanford.edu
\end{flushleft}

\section*{Abstract}
Accurate identification of haplotypes in sequenced human genomes can provide invaluable information about population demography and fine-scale correlations along the genome, thus empowering both population genomic and medical association studies. Yet phasing unrelated individuals remains a challenging problem. Incorporating available data from high throughput sequencing into traditional statistical phasing approaches is a promising avenue to alleviate these issues. We present a novel statistical method that expands on an existing graphical haplotype reconstruction method (shapeIT) to incorporate phasing information from paired-end read data. The algorithm harnesses the haplotype graph information estimated by shapeIT from genotypes across the population and refines haplotype likelihoods for a given individual to be compatible with the sequencing data. Applying the method to HapMap individuals genotyped on the Affymetrix Axiom chip at 7,745,081 SNPs and on a trio sequenced by Complete Genomics, we found that the inclusion of paired end read data significantly improved phasing, with reductions in switch error on the order of 4-15\% against shapeIT across all panels. As expected, the improvements were found to be most significant at sites harboring rare variants; furthermore, we found that longer read sizes and higher throughput translated to greater decreases in switching error, as did higher variance in the size of the insert separating the two reads---suggesting that multi-platform next generation sequencing may be exploited to yield particularly accurate haplotypes. Overall, the phasing improvements afforded by this new method highlight the power of integrating sequencing read information and population genotype data for reconstructing haplotypes in unrelated individuals.
\section*{Author Summary}
Phasing---the reconstruction of haplotypes from diploid genotype data---is a critical analytical step for a large number of downstream applications, ranging from demographic inference to multilocus association studies. Several statistical methods have been developed to reassemble haplotypes from SNP-chip data in unrelated individuals. Yet the advent of sequencing data has added a new dimension to this problem. On the one hand, the use of high coverage data has led to the discovery of new rare variants, which are challenging to phase using traditional methods due to their low frequency in the population. On the other hand, the reads used to generate the sequencing data can shed light on local phasing relationships between neighboring SNPs. In this paper, we present a novel statistical method that exploits this information to generate more accurate haplotypes from next generation sequence data.  

\section*{Introduction}
\noindent Haplotype information is critical to many analyses in population genetics. Many methods of demographic inference rely on haplotype diversity to gain insight into the history of a population, such as possible bottlenecks, range expansions, and strong selection events \cite{leblois, anderson, HCN}. In admixed populations, phasing can be coupled with local ancestry deconvolution to yield estimates of ancestry track length in every individual; these estimates in turn can be used to make inferences on the admixture history of these populations, such as the time and the dynamics of the admixture event \cite{pool, abra, simon}. 

Related individuals share long haplotypes, which can be used to perform long-range phasing. Unrelated individuals also share haplotypes, but these are generally shorter and more difficult to infer. Statistical methods relying on linkage disequilibrium (LD) patterns can be used to infer short range haplotypes, but provide little to no information on longer range phasing \cite{decode}. To date, several algorithms have been developed to perform phasing on large-scale datasets, and these generally appear to yield similar results on commercial arrays (up to 1 million SNPs), with an average 5-7\% error rate between adjacent heterozygous sites for sufficiently large sample sizes \cite{phase, fastphase, beagle}. 

We are interested in exploiting the high coverage afforded by ultra-high throughput sequencing (UHTS) technologies to reduce those error rates in the context of Whole Genome Sequencing (WGS) and Genotyping by Sequencing (GBS). Current algorithms are likely to experience mixed results with such datasets. On the one hand, the increased marker density should lead to the discovery of markers that are on average more tightly linked to one another than those found on ordinary SNP-chips.  On the other hand, many of these markers are expected to be rare variants \cite{tgp}, and these are not easily tagged by common alleles. These can pose a substantial challenge to phasing algorithms, since in those cases the minor allele will only be represented in a handful of individuals, reducing statistical power. 

The gradual shift from SNP-chips to sequencing has spurred a considerable interest in methods designed to physically phase haplotypes from sequencing data. Several statistical algorithms have been developed towards the haplotype assembly problem, whereby haplotypes are directly constructed from the DNA fragments used to assemble the genome. These can be fairly straightforward methods, handling gapped fragments and sequencing errors in a primarily heuristic fashion \cite{fasthare, venter}; others are more statistically involved (such as HASH, which uses an MCMC \cite{HASH} and HapCUT, which relies on a combinatorial graph-cutting algorithm \cite{HapCUT}). In all of these cases, the haplotype reconstruction is done one individual at a time, and is done independently of the population under study. Thus, the success of these algorithms is critically dependent on the sequencing depth and the length of the fragments used to tile the target genome---otherwise, only small haplotype blocks can be reconstructed from the data. 

In addition to the exploration of molecular phasing algorithms, a number of technologies have recently been developed to sequence each chromosome in a cell separately. In a sense, these novel techniques seek to provide a cheaper alternative to sperm typing, in which individual haploid gametes are collected and sequenced. For instance, Kitzman et al. successfully generated the phased genome of a Gujarati individual through the use of fosmid libraries \cite{Shendure}; Fan et al. developed an assay capable of directly haplotyping the genome of single cells\cite{quake}. While these technologies have been shown to reconstruct high-quality haplotypes from diploid cells, they remain comparatively low-throughput, both in terms of cost and running time.  

The goal of this article is to combine population-level LD information with the read data from UHTS to achieve higher accuracy in haplotype reconstruction.  Present technologies generate hundreds of millions of short (30-100bp) reads to cover the target genomes. Some methods (including SOLiD and Illumina) support paired end reads, thus linking nearby regions of the genome (generally 0.3-2kb apart) \cite{solexa, SOLiD}. By default, any read spanning two or more heterozygous sites holds the key to their phasing configuration \cite{Jeff}. Given the massively parallel nature of these technologies, one can expect that a considerable number of sequencing reads will be informative with respect to phase. Thus, incorporating the phasing information from the reads into current population-based algorithms is likely to improve the quality of the inferred haplotypes. In this paper, we present an extension of the popular Delaneau et al. model (implemented by shapeIT)\cite{shapeIT} that accounts for phase-informative reads within single individuals. 

\section*{Results}
\subsection*{Performance assessment on Axiom data}
\paragraph{Cross-population phasing comparison} To assess the performance of the algorithm, we first applied the method to simulation data based on densely genotyped HapMap individuals across 7,745,081 SNPs on the Affymetrix Axiom chip, simulating reads generated by high throughput sequencing. To this end, we generated three equally sized datasets representative of three continental groups: the Yoruba from Ibadan (YRI), CEPH Utah residents with ancestry from northern and western Europe (CEU), and Chinese from Beijing and Japanese from Tokyo (CHB+JPT). Similarly to a number of previous studies, we focused on the X chromosome---genotyped at 156,041 SNPs---generating pseudo-females of known phase by randomly pairing the X chromosomes of the males in each group \cite{fastphase, beagle, shapeIT}. We then simulated reads based on the average characteristics of a standard Illumina run, with a 500bp insert (see methods). These data enabled us to compute switch errors for each population, in the presence and absence of read data (i.e. standard shapeIT). 

\paragraph{Performance benchmarking} We compared the phasing accuracy of {\it seqphase} to that of three other state-of-the-art methods: Beagle \cite{beagle}, fastPHASE \cite{fastphase}, and shapeIT \cite{shapeIT}. Beagle and fastPHASE have been widely used in population genetic studies, the former in great part because of its increased computational efficiency over other methods \cite{beagle, shapeIT}; conversely, shapeIT was run as a control to capture the baseline performance of the algorithm in the absence of reads. We generally found that both shapeIT and {\it seqphase} comfortably outperformed Beagle and fastPHASE in all three datasets, with improvements greater than 20\%. The fact that Beagle performed the worst of the four methods on these panels was expected, given that the algorithm has been shown to perform better in larger panels \cite{beagle, shapeIT}. Furthermore, we found that the inclusion of the simulated paired-end read data led to improvements in switch error rates on the order of 10-15\%; the improvement was found to be highest in the YRI, and lowest in the European and Asian cohorts. 


\paragraph{Switch error and inter-SNP distance}Since we are augmenting the genotype data with reads of a fixed insert size to reduce the error rate of the phasing, we expect to see a depletion of incorrectly phased sites within the distance spanned by the specific insert. To characterize how the inclusion of this information affects the overall distribution of switch errors as a function of inter-SNP distance, we tabulated switch errors by inter-SNP distance both in the presence and absence of reads. To this end, we considered every pair of heterozygous SNPs within 5kb of each other for every individual, and estimated the proportion of those that were correctly phased with respect to one another. Binning these estimates by physical distance and ordering them into ascending order produced the graph in Figure \ref{fig:sevdist}. The black line represents switch errors in the absence of read data, for SNPs separated by 0-5kb. We can see that the switch error steadily increases as a function of physical distance, as expected; for large enough distances, we expect this error to increase asymptotically towards 50\%, as linkage information required for phasing becomes increasingly scarce.  Naturally, there is some variance in these estimates, most likely as a result of the small number of individuals retained in the analysis. The lines in blue and red depict switch errors versus distance for insert sizes averaging 500bp and depths of coverage of 8x and 65x respectively. We can see that, in both cases, there is a depletion in switch errors centered roughly at 500bp, and that this dip is most pronounced at higher depths of coverage. Furthermore, we found that the overall error rate was generally lower at every SNP-SNP distance considered, suggesting that the inclusion of sequencing reads has both a local and a longer range effect on phasing. Finally, we can see that a greater variance in insert size translates to a broader range of SNP-SNP distances---though still centered around 500bp---exhibiting a depletion in switch error rate.

\subsection*{Dependence on sequencing parameters}
We then studied the effect of varying the different sequencing parameters (such as read length, insert size, throughput and insert variance) on overall phasing improvement. To this end, we quantified the effects of altering read length, sequencing throughput and variance of the insert size on the overall switch error, focusing on the subset of JPT pseudo-male individuals used in the previous analysis. In all simulations, the insert size was assumed to be 1kb. 

\paragraph{Read length} Read lengths can be highly variable across different sequencing platforms. The reads can be quite short, as is the case with data currently being generate by SOLiD and Complete Genomics (30 and 35bp, respectively); at the other end of the spectrum, reads generated by 454 technologies can average lengths of 500-700bp, though this increase comes at the expense of throughput. To assess the effects of varying read lengths on the accuracy of the phasing generated by the algorithm, we ran the analysis for reads ranging from 25-200bp in length. As expected, switch error was found to decrease with increasing read lengths, with 200bp reads yielding switch errors that were on average 8\% lower than those obtained with 30bp fragments. 

\paragraph{Throughput} Throughput is expected to be a critical factor in the improvement of inferred haplotype quality, as a larger coverage of the genome should translate to a greater proportion of heterozygous site pairs being represented in the paired end data. To quantify this effect, we re-ran the simulations for throughputs ranging from 0.25-200 Gb. Again, we found that higher throughputs significantly decreased the overall switch error rate in the phased individuals (12\% decrease between 0.25Gb and 200Gb throughputs). Interestingly, simulations run with throughputs greater than 125Gb did not appear to yield significantly different results. This observation suggests that, past a certain level of coverage, most pairs of heterozygous sites within a distance set by the sequencing insert may be represented in the sequencing data. It is thus natural to wonder whether an increased variance in insert size could further help reduce the observed switch error rate. 

\paragraph{Variance in insert size} Sequencing libraries generally exhibit some variation in the size of the insert. This variance traces its way back to the library purification stage: the library is run on a gel, and a region of the gel roughly corresponding to the desired insert size is excised. From a phasing perspective, this variance would be expected to have a favorable effect on the haplotype inference accuracy, as more variation would translate to a broader spectrum of heterozygous site pairs being represented in the sequencing data. To test this hypothesis, we estimated the overall switch error for insert size variances ranging from  0-200bp. We thus found a decreasing, roughly linear relationship between switch error and insert size variance ($R^2=0.98$). 

\subsection*{Phasing the Complete Genomics panel}
Complete Genomics (CG) has released a panel of 60 genomes, taken from several different populations. The CG platform effectively generates paired end data using very short ends (35 bp each) and relatively short inserts (on the order of 300-500bp). To assess the performance of our method on empirical sequencing data, we considered a trio of YRI individuals included in the panel, originally sampled as part of the HapMap project. With the exception of sites that are heterozygous in the child and in both parents, all genotypes in the trio can be phased without error, up to single generation recombination events that have occurred in that generation. In this study, we focused on chromosome 22 for computational ease. As a first pass for this analysis, we opted to exclude {\it singletons}---alleles that appear only once in the entire reference panel. By default, these alleles cannot be phased, since no haplotyping information can be derived from other individuals in the population, all of which are homozygous for the major allele.

\paragraph{Choice of reference panels} An important consideration in this study lies in the construction of the phasing panel. Ideally, for a non-admixed population, one would use a large number of individuals drawn solely from that population (in this case, a large YRI population). However, the CG panel only contains 9 unrelated YRI individuals; thus, the homogeneity in the reference population would come at the expense of statistical power. To determine the optimal reference panel, we thus constructed three reference panels: the set of 9 YRI individuals only, the set of 17 African individuals (comprised of the YRI, Luhya (LWK) and Maasai (MKK) individuals sampled from the HapMap 3 project), and finally the full set of Complete Genomics sequences. Related individuals (such as those in the CEU extended pedigree) were excluded from all phasing panels. A comparison of the performance of our algorithm against shapeIT, fastPHASE and Beagle is shown in Table 3. From Table 3, we can see that that using the full panel of 48 unrelated individuals yields more accurate phasing than focusing on more homogeneous subpopulations for our reference panel. This result was found to hold for all tested algorithms. In the absence of read data, we found that shapeIT outperformed both Beagle and fastPHASE. In particular, Beagle performed the worst of all methods, though the overall gap in performance narrowed considerably with the inclusion of more samples: on average, shapeIT yielded a 63.1\% improvement over Beagle in the full set of individuals, versus a 73.4\% improvement on the YRI-only dataset. With the inclusion of paired-end read data, we found that {\it seqphase} yielded significant improvements in accuracy over shapeIT, though these were more pronounced in the smaller panels. Thus, we observed a 12.9\% reduction in switch error rate using the YRI alone as a reference panel, versus a smaller 3.9\% reduction in the error rate on the full panel. This modest reduction in the number of haplotyping errors is likely a reflection of the low phasing information content of the CG reads. 
 
\paragraph{Phasing rare variants} While the previous results provide an indication of the relative performance of each of the phasing algorithms on the CG panel, they do not offer much insight into the effect of allele frequencies on the quality of the haplotype reconstruction. In practice, we expect rarer variants to be phased more poorly using these conventional statistical methods, as the reference panel would contain less LD information for those sites. To better characterize the performance of these algorithms around variants with low counts in the reference panel, we stratified the number of observed switch errors based on the relative frequencies of the sites at which they were found to occur. These results are shown in Figure 1 for {\it seqphase}, both in the presence and absence of read data. Two observations can be made from this Figure. First, we can clearly see an inverse relationship between site frequency and switch error rate---with singletons exhibiting an error rate close to 50\% in the absence of sequencing read data, as expected. Second, we can see that the incorporation of read data yielded the largest improvements for those variants observed less frequently in the reference panel. 

\paragraph{Constructing a panel of phase-informed rare variants } The comparisons presented in the previous section were performed in the absence of singletons, whose switch errors using any genotype-based method are on the order of 50\%. Yet discarding these singletons altogether is rather limiting, since they can provide deep insight into the patterns of haplotypic diversity of the population under study; this information in turn could be used to substantially improve the accuracy and the resolution of downstream analyses such as local ancestry deconvolution and demographic inferences. In practice, the sequencing data can occasionally circumvent these limitations, by providing a means of recovering some of the variants that occur at very low frequencies in the reference panel. Specifically, our algorithm can phase some of those variants if they happen to be linked to another heterozygous site by paired end reads. Thus, one could consider augmenting the previous dataset with markers harboring rare variants that fall under this scenario. 

To assess the feasibility of this approach, and to determine the number of potential singletons that can be recovered by our method, we scanned the CG raw read database once more, this time retaining rare variants with at least one physical link to another heterozygous site after filtering for sequencing error. We found that 565 (29.8\%) of the 1897 singletons in NA19238 and 525 (32.2\%) of the 1626 in NA19239 had a physical link to another variant site. We thus incorporated this additional set of 1090 markers into our singleton-free dataset, and repeated the above experiment using the four different algorithms. We found that fastPHASE and Beagle yielded mean error rates of 9.54\% and 11.8\% on this new marker set, corresponding to average increases of 22.6\% and 18.6\% for the two methods respectively. In the case of the shapeIT algorithm, we recorded a 51.9\% increase in the observed switch error rate. Conversely, {\it seqphase} yielded an average 3.79\% switch error (an 6.5\% increase). This modest increase is likely due to the incorporation of reads linking singletons to other rare variants that are also difficult to phase. Thus, our algorithm enabled us to incorporate about a third of all singletons in each individual at a minimal cost to the overall phasing accuracy.  The impact of this approach on overall phasing would likely be lower in a larger and more homogeneous population. 

\section*{Discussion}
In this study, we have presented a novel statistical method that integrates sequencing read information and population-level genotype data into a unified framework for haplotype inference. Local phase information contained in reads spanning 2 or more heterozygous sites is used to guide the reconstruction of haplotypes from the other individuals in the sample, by constraining the search space of possible solutions. 

Our simulations on the Affymetrix Axiom chip data have shed a number of insights on the performance of our algorithm against existing phasing methods. Comparing the different methods in the absence of sequencing read data, we found that the lowest error rates were achieved by shapeIT, followed by fastPHASE and lastly Beagle. These results are in line with the results of Delaneau et al. (\cite{shapeIT}), which were established using considerably larger populations for phasing. Based on the same dataset, we found that incorporating sequencing read data (with lengths, insert sizes and throughputs expected of a standard Illumina HiSeq 1000 run) led to significant improvements in the quality of the phasing.  
As expected, these improvements are highly dependent on the sequencing parameters: longer reads, higher throughput and greater variance in the size of the library insert all led to significant decreases in the switch error. Based on these observations, one could postulate that a sequencing approach that would combine high throughput (such as that seen on Illumina, SOLiD, or CG platforms) with longer reads (such as Roche's 454, which now touts average read sizes of 700bp) would lead to even greater improvements. While no such technology exists at the present time, such results may still be achieved in the context of multi-platform next generation sequencing, whereby multiple technologies are harnessed towards a common project. 

In evaluating our method on a Yoruba (YRI) trio of the Complete Genomics dataset, our goal was first to assess the performance of different phasing algorithms on high coverage sequence data. An important challenge facing these statistical haplotypers lies in the extensive presence of rare variants that have generally not been observed on traditional genotyping platforms. At the very end of the site frequency spectrum, the presence of {\it singletons} is particularly problematic, for these sites only appear once in the entire phasing panel, and hence cannot be accurately haplotyped using genotype information alone. As expected, we found that the inclusion of sequencing read data allows for significantly more accurate phasing of singletons. This is most likely because these reads establish a physical link between singletons and nearby more common heterozygous sites. More generally, we found that the phasing improvements afforded by sequencing reads are most significant in rare variants. This observation illustrates the power of our new phasing method in resolving the haplotype configurations of those sites that are most problematic for conventional phasers.  

The transition from simulated to empirical reads brings some inherent challenges. First, some of the reads may contain errors, which if unaccounted for can introduce incorrect phase relationships. Second, the Complete Genomics platform uses short reads (35bp each), separated by a relatively short insert (averaging 400bp \cite{CG}); thus, the extent to which these data would improve phasing remained unclear. We found that, using a simple majority call method to filter out potential errors, the inclusion of the Complete Genomics reads led to significant increases in accuracy over shapeIT. The algorithm performed even better in reference panels where SNP-only based methods were expected to do worse (such as in the YRI and African-only sets). It is important to note that the Complete Genomics platform benefits from a lower sequencing error rate, on the order of 1 variant per 100kb. Thus, it is possible that a more involved error-checking method may be necessary to ensure the quality of the haplotypes derived from other technologies. 

Overall, our results on the Axiom and Complete Genomics datasets provide strong evidence that our algorithm effectively incorporates sequencing read information to yield the most accurate haplotypes of all other methods compared here. In addition to its high accuracy over existing state of the art methods, our algorithm harnesses the powerful computational efficiency of shapeIT. The method has been shown to be linear in the number of markers and conditioning states. It has also been shown to run much faster than most other phasing algorithms, Beagle excepted \cite{shapeIT}. Because the incorporation of reads merely constrains the existing search space generated by the basic algorithm and hence does not involve additional computations, the runtime for phasing is roughly the same in the presence and absence of read data. 

\section*{Materials and Methods}
\subsection*{The basic shapeIT algorithm}
\paragraph{Premise and notation} Similarly to several other phasing algorithms (such as fastPHASE, Mach, IMPUTE), shapeIT relies on the Li and Stephens model for haplotype variation in a population \cite{listephens}. Briefly, we assume that every haplotype observed in the population can be represented as a mosaic of small blocks sampled from all of the other haplotypes in the population. This discretized structure is modeled as a Hidden Markov Model in which each hidden state corresponds to a sampled haplotype. Transitions (or jumps) from one haplotype to the next are modeled as Poisson processes, and hence the arrivals of jump events along the chromosome are exponentially distributed, with parameter $\lambda_m=-4N_er_m/K$ , where $N_e$, $r_m$ and $K$ correspond to the effective population size, the recombination rate between markers $m$ and $m+1$, and the total number of observed haplotypes in the population respectively. The probability of a recombination event in the model is thus given by:
\begin{equation*}
\rho_m=1-e^{-4N_er_m/K}
\end{equation*}
Every iteration of the method implemented in shapeIT proceeds in two general steps. First, the full haplotype space is collapsed into segments comprised of a smaller number of distinct haplotypes.  Second, this set of collapsed states is used as a framework for a Gibbs sampling algorithm, in which the algorithm samples pairs of haplotypes that are compatible with the genotypes of a given individual. 

\paragraph{Haplotype segmentation and the Compact Hidden Markov Model (CHMM)} In the segmentation step, the set {\bf H} of all haplotypes observed in the population is collapsed into disjoint segments of markers, across which the number of distinct haplotypes does not exceed a given threshold $J$; the union of these segments of collapsed states is denoted as {\bf Hg}. This step constrains the complexity of the algorithm to be linear in the number of states $J$, and hence to be independent of the total number of haplotypes in the population; this feature makes the algorithm highly tractable to large-scale datasets. Once the framework of haplotypes is defined, the sampling process follows the Li and Stephens (2003) model of haplotypic diversity in the population. Briefly, the model treats every chromosome as a mosaic of ancestral haplotype blocks. This discretized structure is modeled as a Hidden Markov Model (in this case, a Compact Hidden Markov Model, or CHMM), with J hidden states. Transitions (or jumps) from one haplotype to the next are modeled as Poisson processes, and hence the arrivals of jump events along the chromosome are exponentially distributed. Specifically:
\begin{equation*}
p(z_{m+1}=k_{m+1}|z_{m}=k_m)=(1-\rho_m){c(k_m, k_{m+1})\over c(k_m)}+\rho_m{c(k_m)\over K}
\end{equation*}
where $\rho_m$ is the probability of recombination between markers $m$ and $m+1$ as defined in Li and Stephens 2003, $K$ is the total number of haplotypes in {\bf H}, and $c(k_m)$ is the number of haplotypes in H represented in the collapsed state $k_m$ of {\bf Hg}. 

\paragraph{Gibbs sampling} In the Gibbs sampling step, individual haplotypes are sampled from the CHMM; specifically, a forward-backward-expectation algorithm is executed to compute the posterior probability that a given genotype can be explained by the pairing of any two sampled haplotypes $h_1$ and $h_2$ (henceforth referred to as the diplotype $(h_1, h_2)$). For any given individual, the sampled haplotypes are constrained by the genotypes of the sample. While the space {\bf S} of all possible haplotypes compatible with these genotypes is intractably large for even small datasets, this computational limitation can also be circumvented by segmentation: the vector of genotypes is broken down into consecutive segments of fixed size $B$ ambiguous markers (either heterozygous or missing; $B=3$ in the seminal paper), each with 8 possible haplotypes. The aim of the Gibbs sampler is thus to estimate the most likely diplotype transitions between each pair of consecutive segments. This goal is achieved by consecutively sampling adjacent segments in {\bf Sg} based on the forward and backward probabilities derived from the CHMM. At every round of the Gibbs sampler, the algorithm keeps track of the number of observed transitions between two adjacent nodes in the {\bf Sg} graph. 

\paragraph{Phasing} The ultimate step of the algorithm involves phasing the individuals based on the full set of simulated haplotype configurations, and based on the total number of observed transitions across {\bf Sg} for every individual. To minimize the switch error rate between pairs of heterozygous sites, a Viterbi algorithm is applied over the graph {\bf Sg} to determine the maximum-likelihood haplotype configuration, using the empirical transition counts as transition probabilities

\subsection*{Incorporating sequencing data}
\paragraph{Statement of the problem} The inclusion of paired end read data can provide local phasing information between select pairs of heterozygous sites. These sites may be spanned by a single read, or by two separate reads linked by an unsequenced insert. Let $\Omega$ denote the set of all pairs of genotypes phased using the sequencing data, and let $\Omega'$ denote the subset of those phasings that are incompatible with the haplotypes derived from a standard algorithm such as fastPHASE, Beagle or ShapeIT. Assuming for now that there are no errors in the sequencing data, the pairs of heterozygous sites in $\Omega'$ represent actual phasing errors in the algorithm. Our goal is to correct the inferences of the algorithm wherever this information is available. This goal is achieved through a two step modification of the original algorithm: we first implement a constrained version of the Gibbs sampler to sample combinations of heterozygous sites that are consistent both with the sequence data and with haplotypes resolved from the population; we then count {\it multisegment haplotypes} (instead of pairs of segments, as in \cite{shapeIT}) to phase those sites where sequence data are available. 

\paragraph{Constrained Monte Carlo method} We constrain the original (unconstrained) sampler outlined by Delaneau et al based on two factors. First, we exclude haplotypes within a segment that are inconsistent with the read data. For SNPs that are relatively close to one another and for which read data are available, it is possible to immediately exclude haplotypes within a segment, thus constraining the set of possible transitions from the previous segment. Second, we eliminate haplotypes inconsistent with read data pertaining to a segment that has already been sampled. Every time a diplotype is sampled within a segment, the set of compatible diplotypes at subsequent segments is constrained by the read information at that segment (Figure 1). To facilitate mixing of the chain, we randomly alternate between sampling haplotypes along the chromosome from left to right, and from right to left. A similar approach has been implemented in Beagle \cite{beagle}. 

\paragraph{Counting multilocus haplotypes} The original algorithm carries out phasing by counting the most common pair of alleles that arise at adjacent segments in the simulation. However, this approach is no longer feasible in the context of sequence data. By solely focusing on adjacent pairs of segments, we would be ignoring the more complex phasing relationships that can be gleaned from the reads; as a result, it would be impossible to ensure that the final phasing would be faithful to the sequence data. Instead, we count entire haplotypes wherever sequence data are available.To this end, we subdivide the chromosome under consideration into discrete blocks covered by informative reads. Starting from one end of the chromosome, a given block starts at a heterozygous site that is linked to a downstream heterozygous site (but not to any upstream sites); the block ends when we reach a heterozygous site beyond which there are no more downstream links (ie. no links originating inside the block terminate downstream of that site). We then phase each block by selecting the multilocus haplotype that appears most frequently in that block. Naturally, sites for which no sequence data are available are phased using the standard method of Delaneau et al.

\paragraph{Post-hoc versus iterative phasing improvements} It is important to note that, under this implementation, the read data are used to help improve the phasing of the sample in two ways. First, the constraining of the haplotype space during the Gibbs sampling process allows for the post-hoc, independent improvement of haplotype phasing in every individual in the sample; in other words, we first ``fix'' phasing configurations that are determined to be incorrect on the basis of the read data. Second, and perhaps more interestingly, the phasing improvements in a given individual are used to guide the haplotype reconstruction of all other samples in the reference panel. This is because once a sequencing read-compatible haplotype is sampled for a given individual, this haplotype is re-inserted into the CHMM, thus influencing the structure and the posterior probabilities associated with the Markovian framework of the algorithm.

\paragraph{Measures of performance} Traditionally, the benchmark of performance for any phasing algorithm is the switch error \cite{fastphase}, defined as the ratio of incorrectly phased heterozygous sites to the total number of heterozygous sites:
\begin{equation*}
S_{err}={h_{err}\over h_{total}}
\end{equation*}

\subsection*{Simulations on the Axiom data}
\paragraph{Sample preparation} To assess the performance of the algorithm, we considered HapMap individuals genotyped at high density on the Affymetrix Axiom chip. These individuals are grouped into 4 different populations: the Yoruba from Nigeria (YRI),  the CEPH Utah residents with ancestry from northern and western Europe (CEU), the Chinese from Beijing (CHB), and the Japanese from Tokyo (JPT). Pseudo-individuals of known phase were constructed by randomly pairing full X chromosomes (excluding pseudo-autosomal region) from males in each population. These pairings yielded 14 YRI, 14 CEU, 11 CHB and 10 JPT pseudo-females. Since we are interested in comparing the relative improvement in phasing afforded by the inclusion of sequence data in each of the populations, we opted to augment these sets of individuals with additional female X chromosomes. Specifically, we augmented the YRI and CEU sets with 29 females each, yielding two sets of 43 individuals. For the Asian populations, we combined the JPT and CHB simulated females and included another 22 females randomly drawn from a pooled set of JPT and CHB individuals.

\paragraph{Read simulation} To assess the effect of incorporating pared end read data on overall switch error rates and correct haplotype length distributions on the Axiom data, we chose to simulate paired end reads that would be expected from an Illumina Hi-Seq 1000 run. Specifically, we assumed reads 100bp in length each, throughput of 200Gb, and an average insert size of 500bp. To simulate some of the read variability inherent to library construction, we sampled insert sizes for every read from a normal distribution centered at 500bp, and with standard deviation 100bp. Throughout the simulation, we only include those reads that span at least two heterozygous sites. For the analyses assessing performance dependence on sequencing parameters, we opted to use a mean insert size of 1kb, as a representative middle ground between what would be expected using Illumina and SOLiD technologies. 

\subsection*{Preparation of the Complete Genomics data}
A unique feature of the Complete Genomics platform lies in the preparation of the sequencing substrate. Once the target genomic DNA is sheared, the resulting fragments are ligated to adapters, circularized, and then amplified by RCR \cite{rcr}; this step differs from current Illumina and SOLiD platforms, which rely on a PCR amplification step of the entire library \cite{solexa, SOLiD}.  The CG protocol leads to the formation of DNA nano-balls (DNBs)---coils of tandem-repeated single stranded DNA. Each DNB is effectively comprised of two half-DNBs, separated by a mate gap that averages about 350bp. Each half-DNB in turn is comprised of 4 DNA fragments (of lengths 10,10,10 and 5 nt respectively); these fragments may have some overlap, or they may be separated by small gaps (generally $<3$ nt). Thus, half-DNBs are somewhat equivalent to ``reads'' averaging 35bp, though with some variation \cite{CG}. 

\paragraph{Phase-informative DNBs} To identify phase-informative DNBs (or half DNBs), we resorted to alignments  generated by CG using a custom algorithm that accounts for the structure of these reads \cite{CG}. We excluded all half-DNBs that did not map to one and only one genomic location (multiple hits would be useless in the context of phasing). Of those, we then filtered out all half-DNBs that did not span at least one heterozygous site (based on the CG genotype calls), as well as all full-DNBs that did not cover at least 2 heterozygous sites. Finally, any half-DNBs containing missing data or bases inconsistent with respect to the genotype call at those heterozygous sites were discarded as well. 

\paragraph{Error handling} While the previous filtering method is successful in identifying informative reads for phasing, it is unable to screen, it cannot detect sequencing errors that are still consistent with the called genotype (e.g an A allele being miscalled as G at an A/G heterozygous site). Yet these errors can result in DNBs or half-DNBs containing erroneous phase information. To account for these errors, we kept track of all informative 2-locus haplotypes that could be gleaned from the data. We then performed a majority call vote: the highest ranking haplotype by raw count was chosen as the correct haplotype. This was done irrespective of the total number of haplotypes (ie. we did not exclude pairings for which only a small number of reads were available). However, we excluded the read data in cases where more than one haplotype was ranked highest. 

\section*{Acknowledgments}
We thank Jeff Kidd and Simon Gravel for useful discussions and valuable comments on the manuscript.  We also thank Jeanette Schmidt for providing us with the Affymetrix Axiom chip genotype data, and Steve Lincoln for supplying us with the sequencing read data from Complete Genomics.

\bibliography{template}

\section*{Figure Legends}
\paragraph{Figure 1} Overview of the method used to integrate sequencing reads into the population-based phasing. In this example, we consider an individual with three consecutive heterozygous sites. In the absence of any other information, the alleles at these sites can be grouped into 4 possible diplotypes. However, the presence of a sequencing read (shown as green bases) narrows down the list of compatible diplotypes to 2. Note that we must still rely on population-level LD information to identify the most likely of the remaining diplotypes.  

\paragraph{Figure 2} Switch error for {\it seqphase} as a function of physical distance on the YRI Axiom panel for different simulation parameters of sequencing read data. The horizontal green line represents the basal 0\% switch error rate. The arrow highlights the depletion in switch errors observed in the presence of paired end reads at SNP pairs separated by a distance equivalent to the insert size (in this case for an insert of 500bp). 


\paragraph{Figure 3} Dependence of phasing accuracy on (a) read length, (b) throughput, and  (c) variance in the insert size, assuming a mean insert size of 1kb.  

\paragraph{Figure 4} Switch error rate in the CG panel versus the observed number of derived alleles in the full reference panel of 48 unrelated individuals. To reduce the effect of disparities in the number of sites included in each category (e.g singletons occurring more frequently than tripletons), we grouped sites with 3-6, and 7-10 observed derived alleles.

\begin{figure}[p]
\begin{center}
\subfigure[No reads]{
\includegraphics[width=4in]{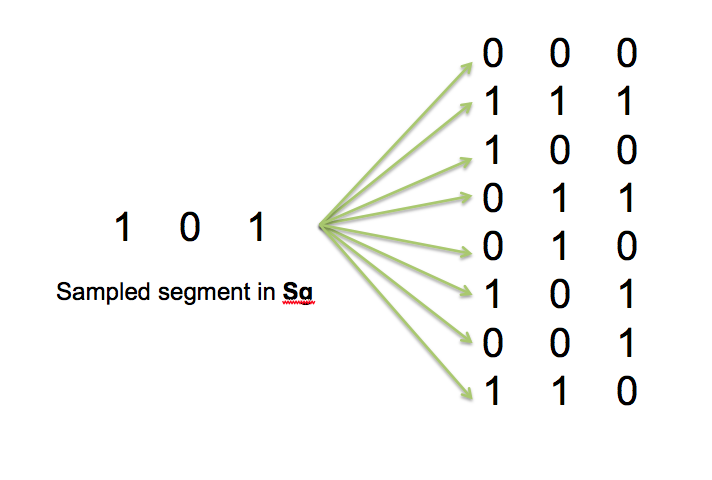}
\label{fig:gplot1}}\\
\subfigure[Reads]{
\includegraphics[width=4in]{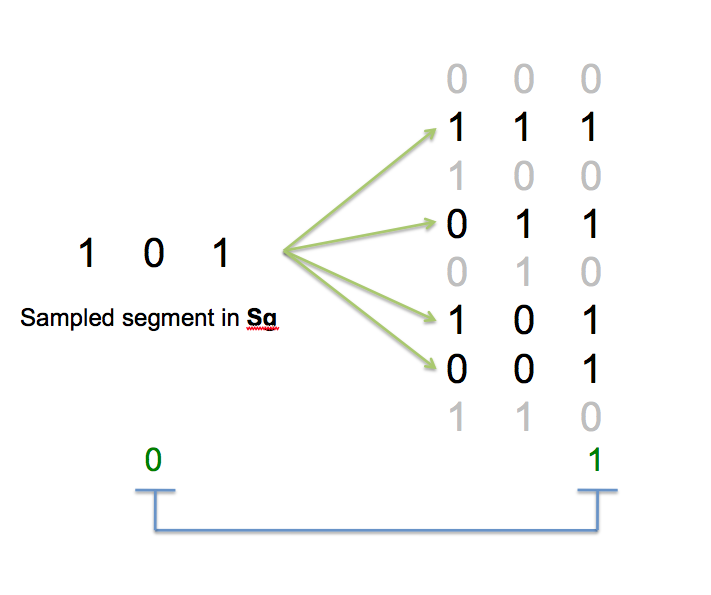}
\label{fig:gplot1}}\\
\end{center}
\caption{
{\bf Constraining the search space using sequencing read data.}  Every phase-informative read halves the number of candidate haplotypes in the Gibbs sampler. 
}
\label{Figure_label}
\end{figure}

\begin{figure}[p]
\begin{center}
\includegraphics[width=0.9\textwidth]{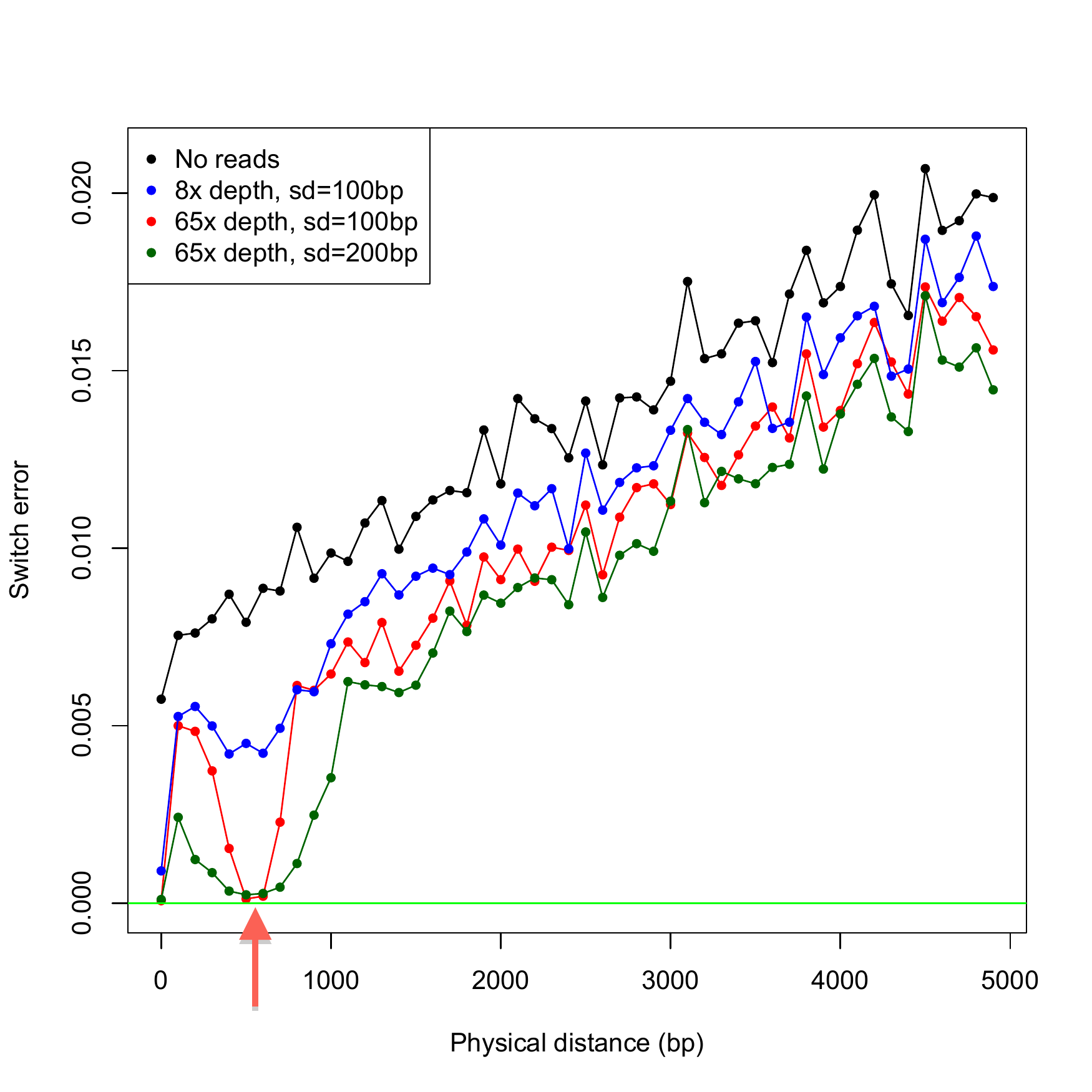}
\end{center}
\caption{
{\bf Switch error versus heterozygous SNP distance in the absence and presence of sequencing reads.}  The incorporation of sequencing data affects both short and long-range phasing. 
}
\label{Figure_label}
\end{figure}

\begin{figure}[p]
\begin{center}
\subfigure[Switch error versus read length]{
\includegraphics [width=0.45\textwidth, height=0.28\textheight]{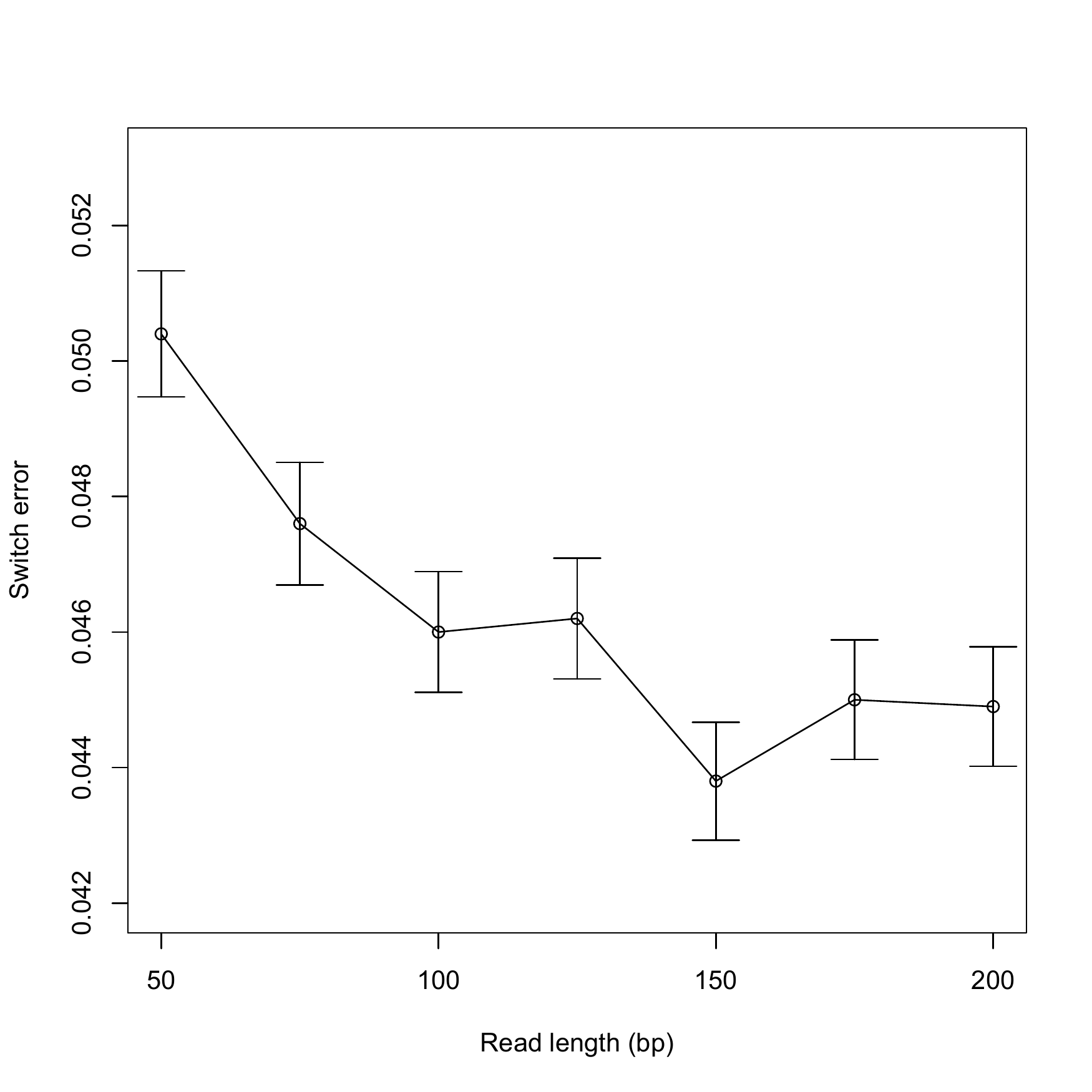}
\label{fig:gplot1}}\\
\subfigure[Switch error vs standard deviation]{
\includegraphics [width=0.45\textwidth, height=0.28\textheight]{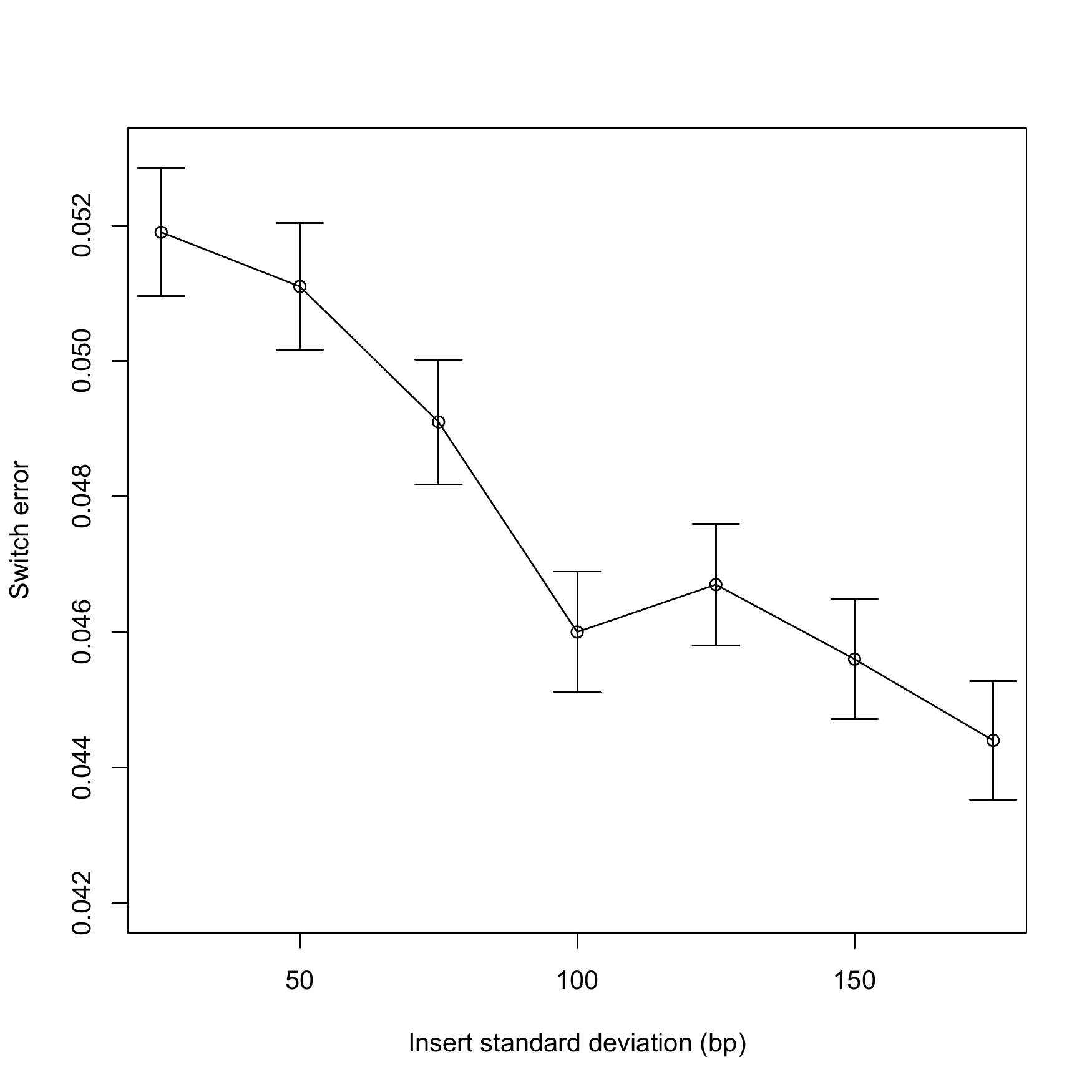}
\label{fig:rplot4}}\\
\subfigure[Switch error vs throughput]{
\includegraphics [width=0.45\textwidth, height=0.27\textheight]{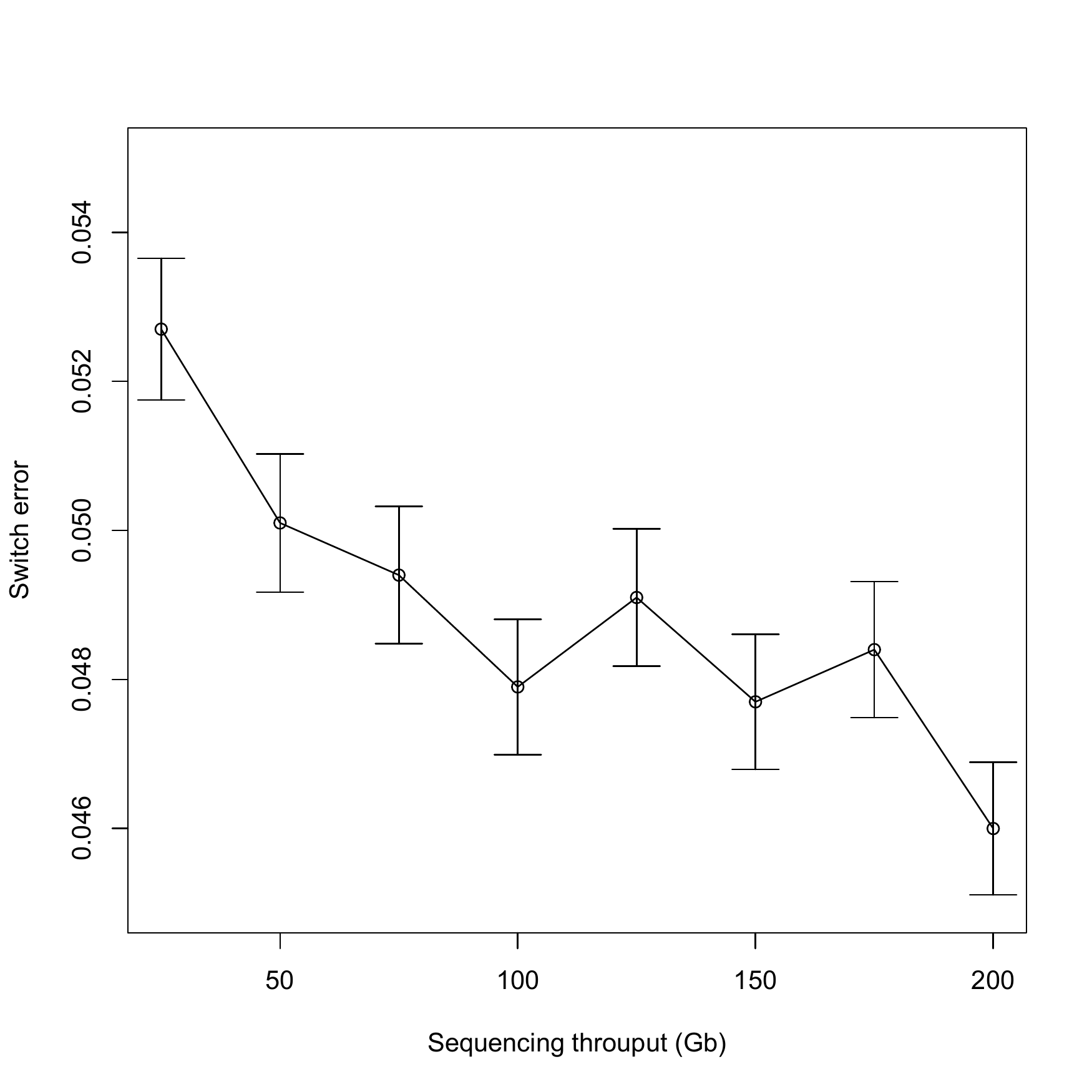}
\label{fig:rplot4}}
\end{center}
\caption{
{\bf Dependence of phasing accuracy on different sequencing parameters.}  Longer reads, larger variances in insert size and greater coverage all lead to significantly lower switch error rates. 
}
\label{Figure_label}
\end{figure}

\begin{figure}[p]
\begin{center}
\includegraphics[width=0.9\textwidth]{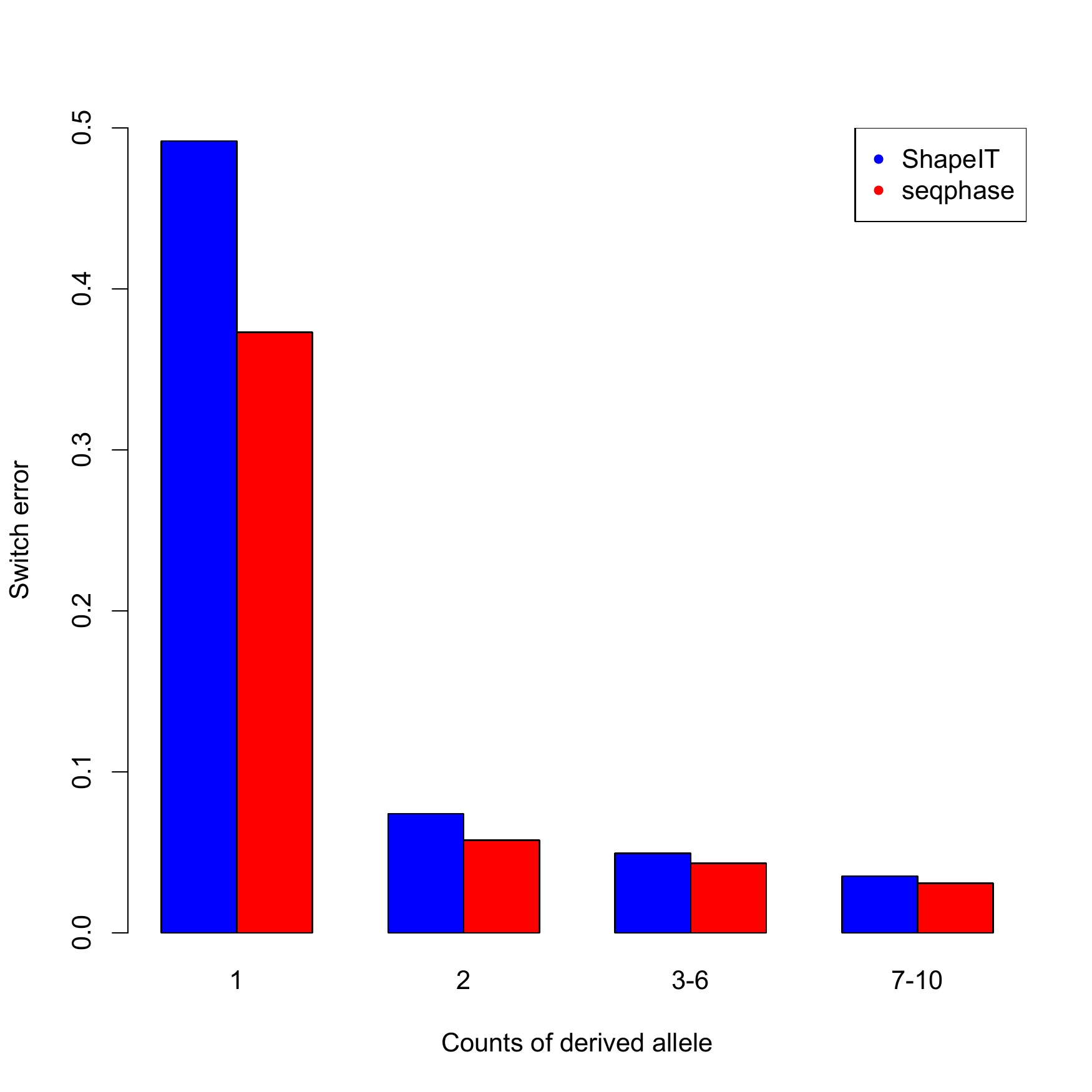}
\end{center}
\caption{
{\bf Switch error versus derived allele frequency in the Complete Genomics panel.}  The incorporation of sequencing data leads to the highest improvements in rare variants.  
}
\label{Figure_label}
\end{figure}


\newpage
\section*{Tables}
\begin{table}[!ht]
\centering
\caption{\bf{Switch error rates on the Axiom genotype data with simulated reads}}
\begin{tabular}{| c || c ||c| c |c|c|}
\hline
Population& Het. markers& Beagle& fastPHASE& shapeIT&{\it seqphase} \\
\hline
YRI&113,847&3.90\%&3.10\%&2.05\%&1.73\%\\
&&$(-55.6\%)$&$(-44.2\%)$&$(-15.2\%)$&\\
\hline
CEU&86,870&2.88\%&2.55\%&1.86\%&1.66\%\\
&&$(-42.4\%)$&$(-34.9\%)$&$(-10.8\%)$&\\
\hline
CHB+JPT&79,391&2.82\%&2.47\%&2.25\%&2.00\%\\
&&$(-29.1\%)$& $(-19.0\%)$&$(-11.1\%)$&\\
\hline
\end{tabular}
\begin{flushleft}Switch error rates for 3 balanced Axiom populations (43 individuals each) using fastPHASE, Beagle, shapeIT and {\it seqphase} with reads of 100bp, 500bp inserts, insert size standard deviation of 100bp. Numbers in parentheses represent \% decrease in switch error when using {\it seqphase}
\end{flushleft}
\label{tab:label}
 \end{table}
 
\begin{table}[!ht]
\centering
\caption{
\bf{Switch error rates on the Complete Genomics dataset}}
\begin{tabular}{|c|c||c|c|c|c|}
\hline
Reference&Size&Beagle&fastPHASE&ShapeIT&{\it seqphase}\\
\hline
YRI &9&33.2\%&15.4\%&8.82\%&7.68\%\\
\hline
YRI+LWK+MKK &17&22.6\%&9.94\%&5.47\%&5.06\%\\
\hline
Full panel&48&9.96\%&7.80\%&3.68\%&3.56\%\\
\hline
\end{tabular}
\begin{flushleft}
Switch error rates for Beagle, fastPHASE, shapeIT and {\it seqphase} on the Complete Genomics Yoruba trio, benchmarked on three possible reference panels (Yoruba only, African only, and full set of unrelated individuals). Singletons were excluded from all reference panels. 
\end{flushleft}
\label{tab:label}
 \end{table}

\begin{table}[!ht]
\centering
\caption{
\bf{Incorporation of phase-informed rare variants}}
\begin{tabular}{|c||c|c|}
\hline
&No singletons&Read-linked\\
\hline
Beagle&9.96\%&11.80\%\\
\hline
fastPHASE&7.80\%&9.55\%\\
\hline
shapeIT&3.68\%&5.59\%\\
\hline
{\it seqphase}&3.56\%&3.79\%\\
\hline
\end{tabular}
\begin{flushleft}
Switch error rates for Beagle, fastPHASE, shapeIT and {\it seqphase} on the Complete Genomics Yoruba trio, both in the absence of singletons, and in the presence of singletons linked to other heterozygous sites by one or more sequencing reads. 
\end{flushleft}
\label{tab:label}
 \end{table}

\end{document}